\documentclass[a4paper]{spie}  %>>> use this instead for A4 paper
\usepackage[]{graphicx}

\title{Elevation angle dependence of the SMA antenna focus position}

\author{Satoki Matsushita\supit{a,c}, Masao Saito\supit{b,c},
	Kazushi Sakamoto\supit{b,c}, Todd R. Hunter\supit{c},\\
	Nimesh A. Patel\supit{c}, Tirupati K. Sridharan\supit{c},
	and Robert W. Wilson\supit{c}
\skiplinehalf
\supit{a}Academia Sinica Institute of Astronomy and Astrophysics,\\
	P.O. Box 23-141, Taipei 106, Taiwan, R.O.C.; \\
\supit{b}National Astronomical Observatory of Japan,
	Mitaka, Tokyo 181-8588, Japan; \\
\supit{c}Harvard-Smithsonian Center for Astrophysics,\\
	60 Garden Street, Cambridge, MA 02138, USA
}

\authorinfo{Further author information: (Send correspondence to S.M.)\\
	S.M.: E-mail: satoki@asiaa.sinica.edu.tw}
\begin{document}
  \maketitle

%%%%%%%%%%%%%%%%%%%%%%%%%%%%%%%%%%%%%%%%%%%%%%%%%%%%%%%%%%%%%
\begin{abstract}
We report the measurement results and compensation of the antenna
elevation angle dependences of the Sub-millimeter Array (SMA)
antenna characteristics.
Without optimizing the subreflector (focus) positions as a function
of the antenna elevation angle, antenna beam patterns show lopsided
sidelobes, and antenna efficiencies show degradations.
The sidelobe level increases and the antenna efficiencies decrease
about $1\%$ and a few \%, respectively, for every $10^{\circ}$ change
in the elevation angle at the measured frequency of 237~GHz.
%At higher frequencies, the sidelobe level and the antenna
%efficiencies are expected to be worse significantly, which will
%make high frequency observations very difficult.
We therefore obtained the optimized subreflector positions for X
(azimuth), Y (elevation), and Z (radio optics) focus axes at various
elevation angles for all the eight SMA antennas.
The X axis position does not depend on the elevation angle.
The Y and Z axes positions depend on the elevation angles, and are
well fitted with a simple function for each axis with including
a gravity term (cosine and sine of elevation, respectively).
In the optimized subreflector positions, the antenna beam patterns
show low level symmetric sidelobe of at most a few\%, and the antenna
efficiencies stay constant at any antenna elevation angles.
Using one set of fitted functions for all antennas, the SMA is now
operating with real-time focusing, and showing constant antenna
characteristics at any given elevation angle.
\end{abstract}

%>>>> Include a list of keywords after the abstract


%%%%%%%%%%%%%%%%%%%%%%%%%%%%%%%%%%%%%%%%%%%%%%%%%%%%%%%%%%%%%
\section{INTRODUCTION}
\label{sect:intro}  % \label{} allows reference to this section

\begin{figure}
   \begin{center}
   \begin{tabular}{c}
   \includegraphics[height=7cm]{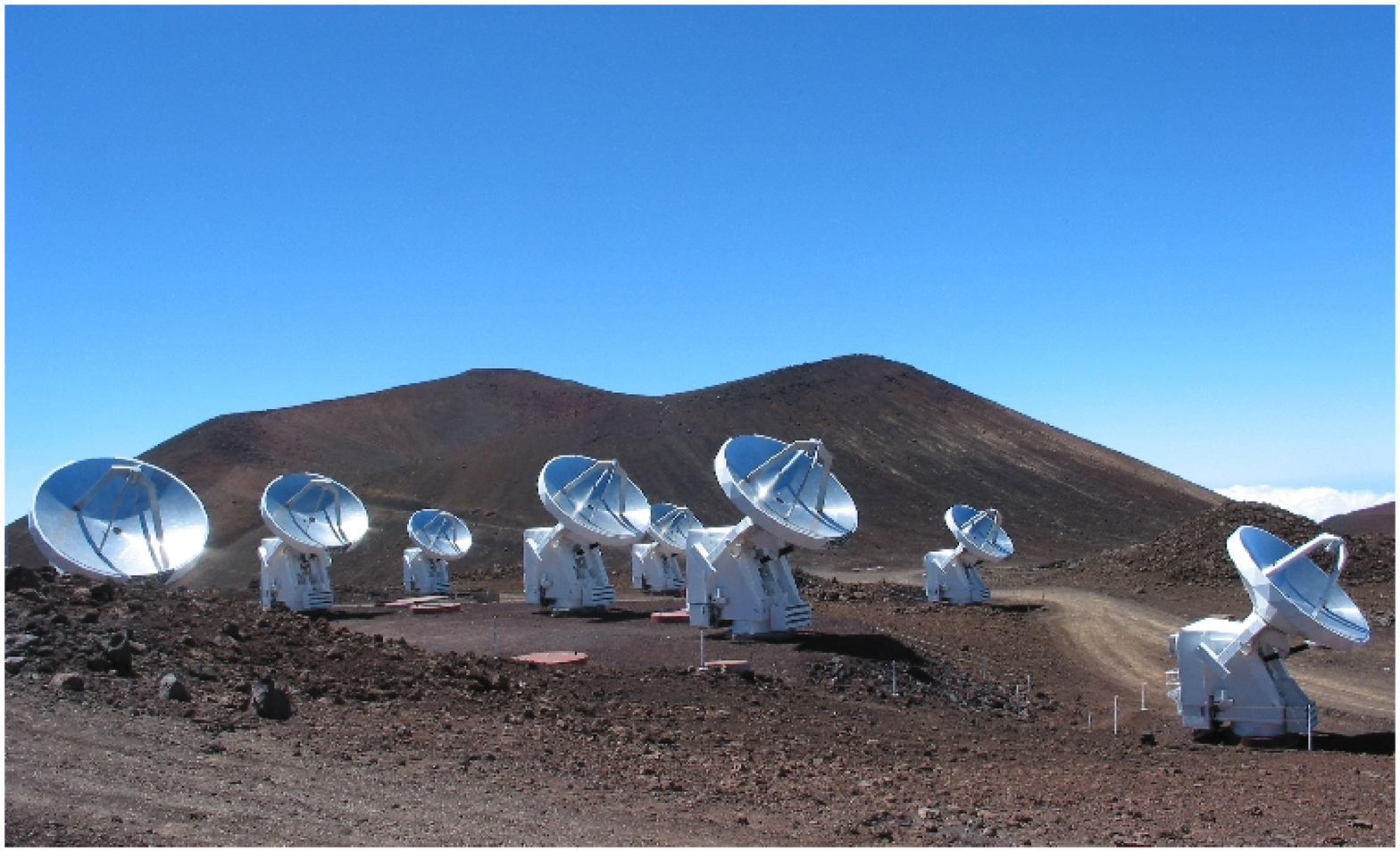}
   \includegraphics[height=7cm]{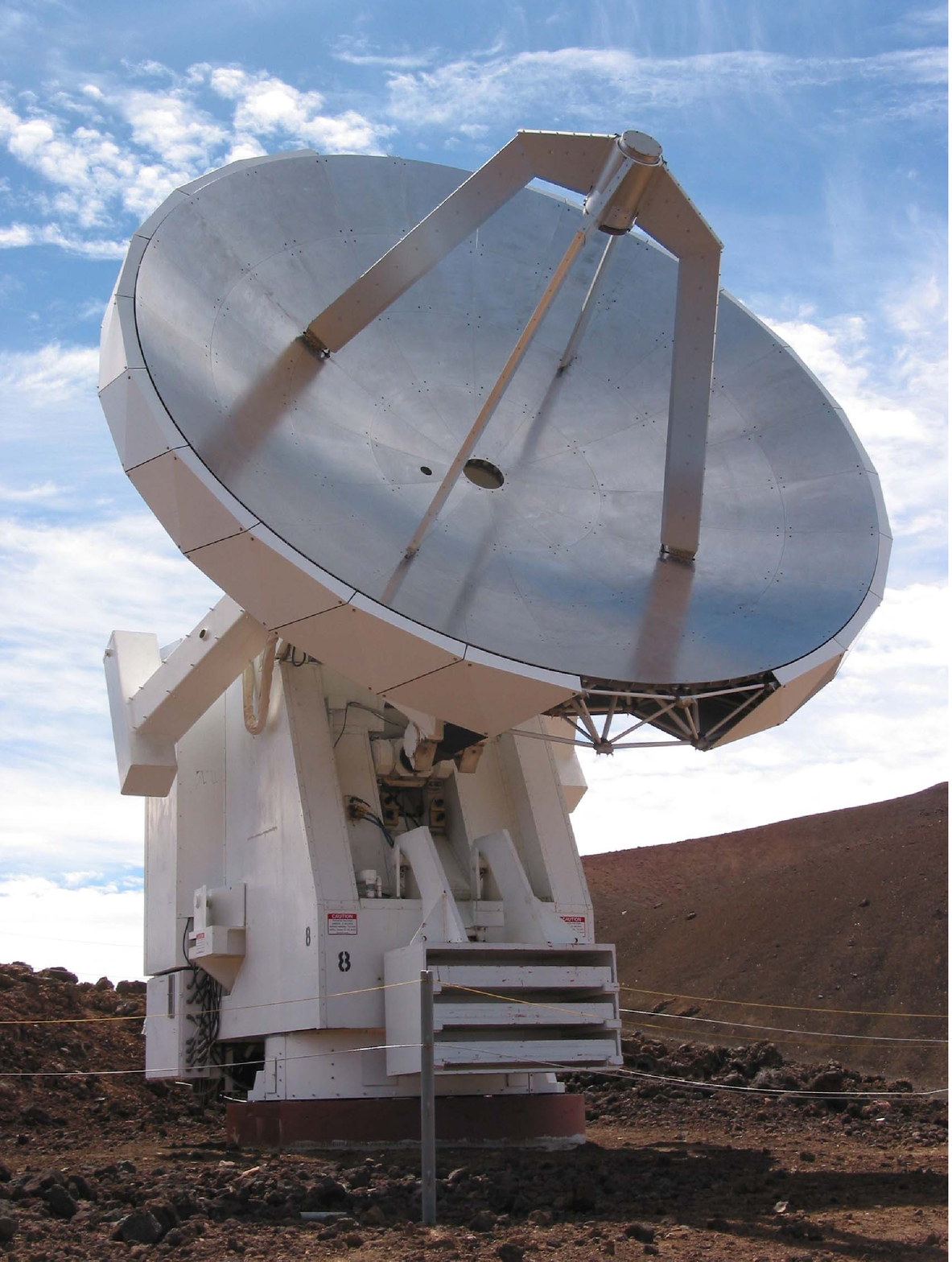}
   \end{tabular}
   \end{center}
   \caption[sma]
%>>>> use \label inside caption to get Fig. number with \ref{}
   {\label{fig:sma}
    Pictures of the Submillimeter Array (SMA; left) and an SMA
    antenna (right).
    The SMA consists of eight 6~m diameter antennas.
    The main reflector consists of 72 aluminum panels supported by
    carbon fiber tubes and steel nodes.  The backup structure is
    usually covered, but in this figure, the bottom part is not
    covered yet, so that the backup structure is visible.
    The subreflector is supported by a quadrupod.}
   \end{figure}

The Submillimeter Array (SMA; left figure of Fig.~\ref{fig:sma}) is
the world's first dedicated submillimeter interferometer, and
consists of eight 6~m antennas.
It will cover the frequency range of 180--900~GHz with a 2~GHz
bandwidth in both upper and lower side bands.
An SMA antenna (right figure of Fig.~\ref{fig:sma}) has a 6~m
diameter main reflector, and its surface consists of 72 machined
cast aluminum panels.
These panels are supported by an open backup structure consists of
carbon fiber tubes and steel nodes \cite{ho04}.
The secondary reflector (subreflector) is supported by a quadrupod.
The main reflector and the quadrupod are deformed by gravity as the
elevation angle of the telescope change \cite{raf91}.
%Since the main reflector consists of many parts,
%The main reflector can be deformed by the gravity, and the degree of
%the gravitational deformation can be changed with the elevation angle
%of the main reflector.
%In addition, the secondary reflector (subreflector) is supported by
%a quadrupod, and the position of the sub-reflector can also be
%changed by the gravity \cite{raf91}.

%%  Use following command to specify that graphics file is in 
%%  a directory other than this LaTeX source file.
%%  Note use of / to separate subdirectories, for UNIX and Windows OS.
%%\graphicspath{{H:/HANSON/SPIESTY/}}
%-------------
%-------------

The gravitational deformation of the main reflector and the position
change of the subreflector cause changes of the antenna beam shapes
(beam patterns) and the focus positions, and affect the quality of
the observational data, especially high frequency observations,
extended object imagings, and mosaicing.
It is therefore important to figure out the effect of the gravity on
the SMA antenna reflectors, and to find out the ways to avoid or
compensate these effects.
We performed antenna beam pattern and efficiency measurements along
various telescope elevation angles (Sect.~\ref{sect:measfix}).
We then measured the optimized focus positions at various telescope
elevation angles, and modeled the focus curves with a simple
function (Sect.~\ref{sect:opt}).
The optimized focus positions are verified with the measurements of
beam pattern and efficiencies using the derived focus curves
(Sect.~\ref{sect:measopt}).
Finally, we apply these results to the real-time subreflector
position optimization (Sect.~\ref{sect:real}).

%%%%%%%%%%%%%%%%%%%%%%%%%%%%%%%%%%%%%%%%%%%%%%%%%%%%%%%%%%%%%
\section{MEASUREMENT RESULTS WITH FIXED FOCUS POSITION}
\label{sect:measfix}

At the early days of the SMA, the subreflector positions were
optimized at the elevation angle of $\sim19^{\circ}$.
This is due to the setup of holography measurements to adjust
the SMA antenna main reflector surfaces.
A 232~GHz radiation source (beacon) for the holography measurements
is located at the catwalk of the Subaru Telescope, which is located
at north-east of the SMA.
The near-field focus position is optimized and the surface of a main
reflector is adjusted by the holography measurements at an elevation
angle determined by the elevation angle of the beacon from the pad on
which the antenna was located when the measurements were made, which
is about $19^{\circ}$ from the SMA inner pads \cite{sri02,sri04}.
The far-field focus position is estimated from this near-field focus
position, and therefore the far-field focus position is also
optimized at the elevation angle of $\sim19^{\circ}$.
Furthermore, the subreflector position was fixed at this position in
whatever antenna elevation angles.
Since the focus position changes with the antenna elevation angle
due to the gravitational deformation as mentioned above, antenna beam
patterns can be deformed and antenna efficiencies can be degraded
as a function of the antenna elevation angle.

%%%%%Sometimes it is necessary to precede the double slash 
%%%%%by \verb|\protect| to get the desired result, 
%%%%%for example, in article titles.

%%-----------------------------------------------------------
\subsection{Beam Pattern Maps with Fixed Focus Position}
\label{sect:bpfix}

\begin{figure}
   \begin{center}
   \begin{tabular}{ccc}
   \includegraphics[height=5.1cm]{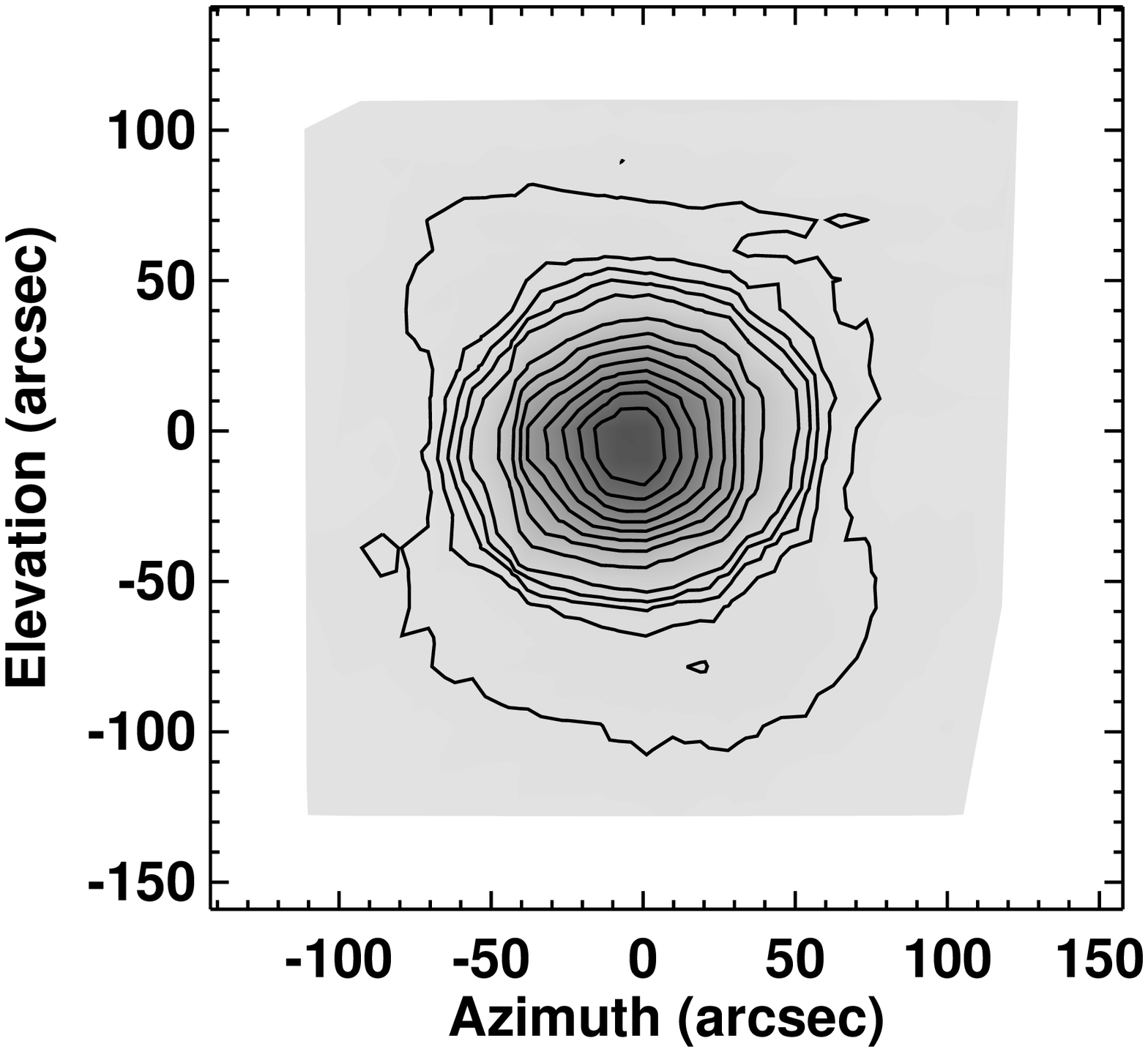}
   \includegraphics[height=5.1cm]{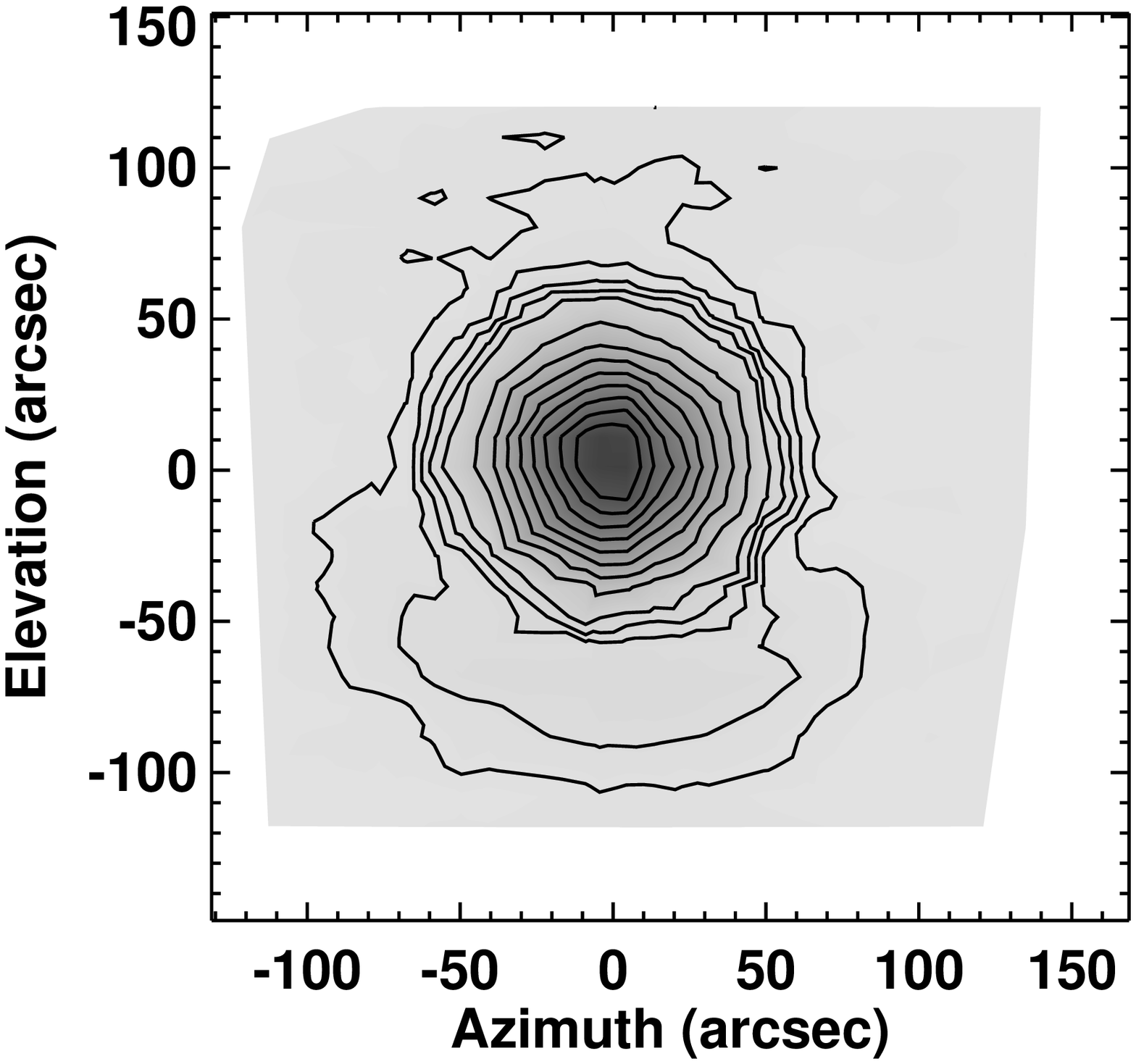}
   \includegraphics[height=5.1cm]{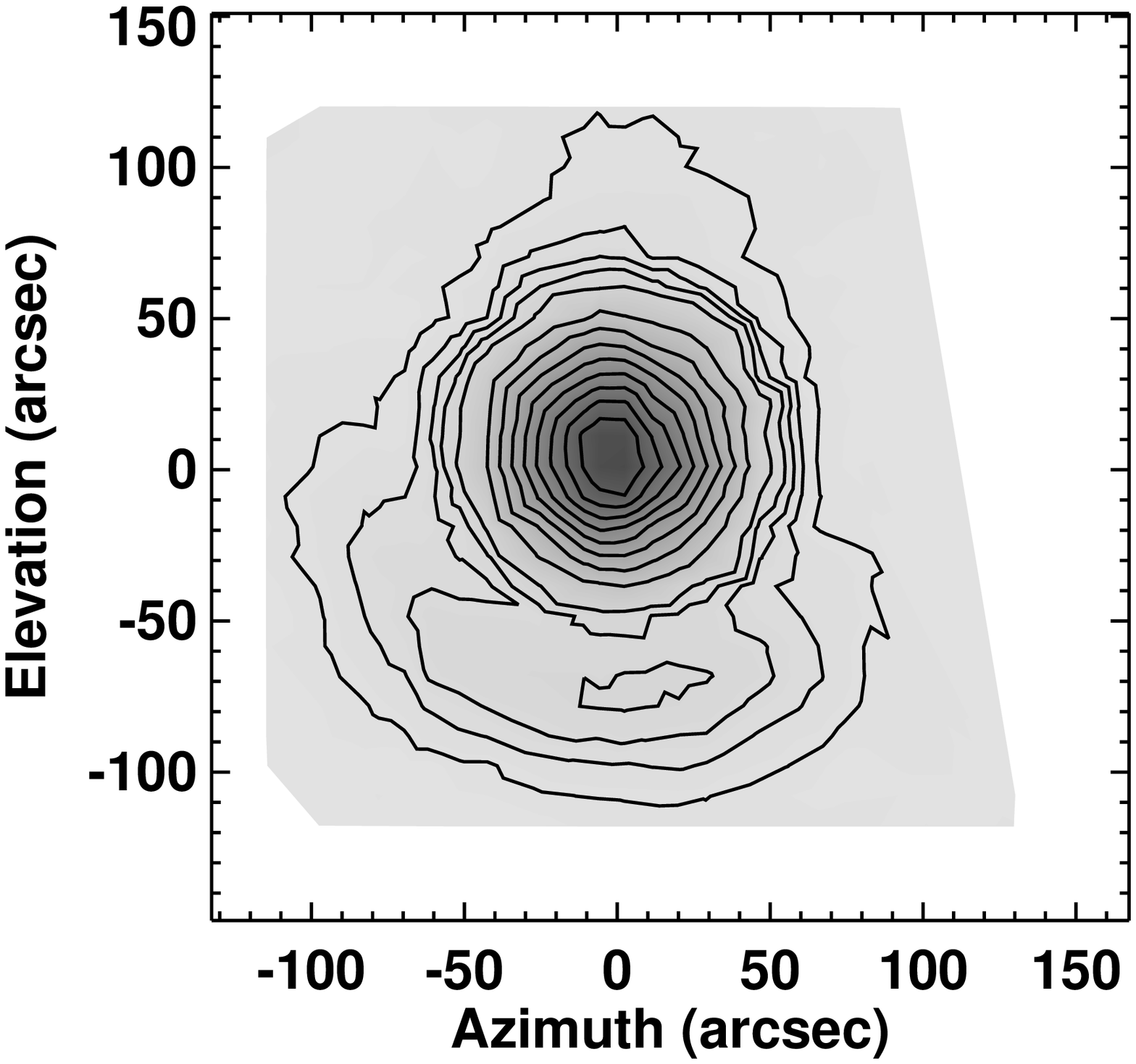}\\
   \includegraphics[height=5.3cm]{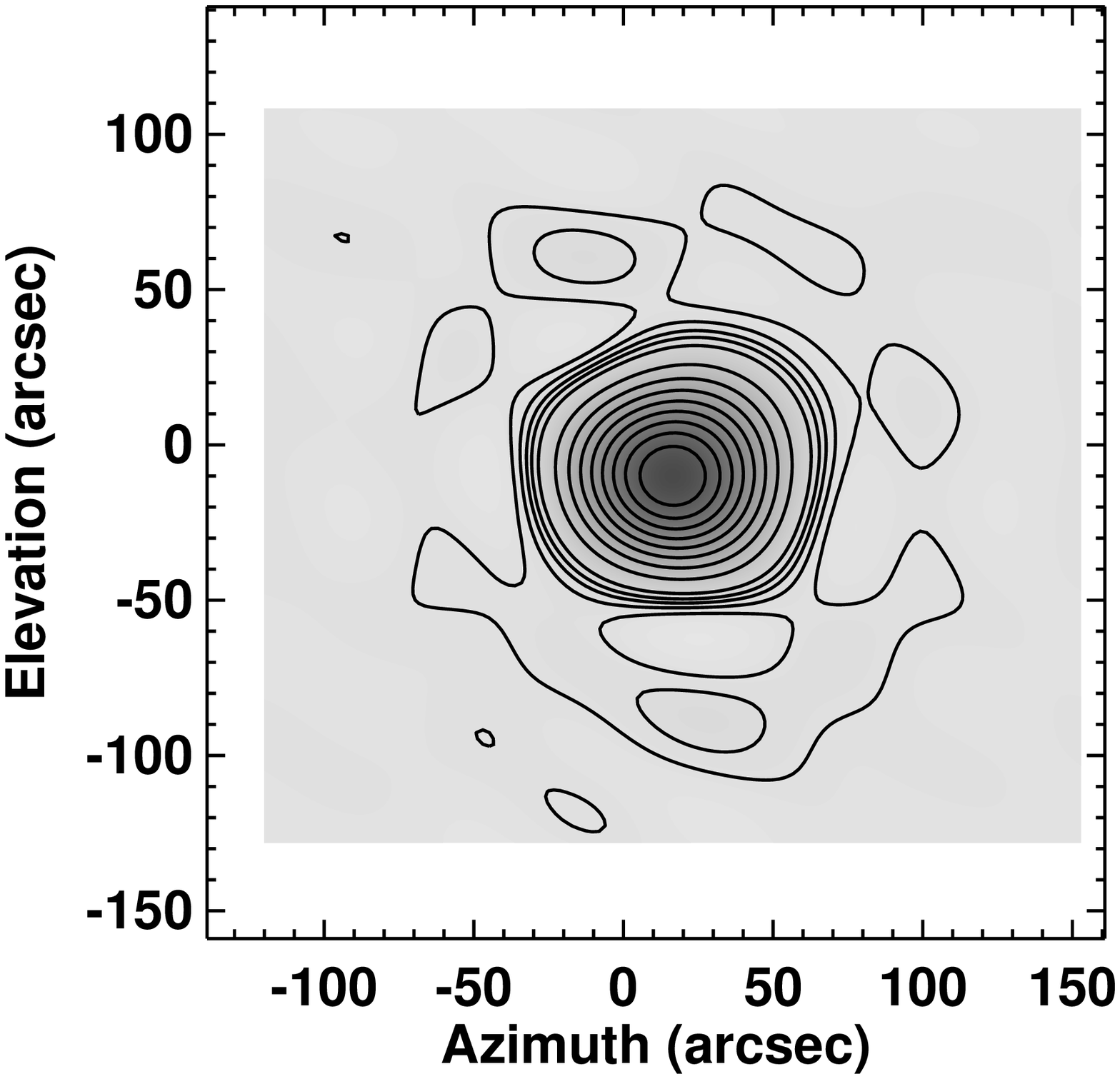}
   \includegraphics[height=5.3cm]{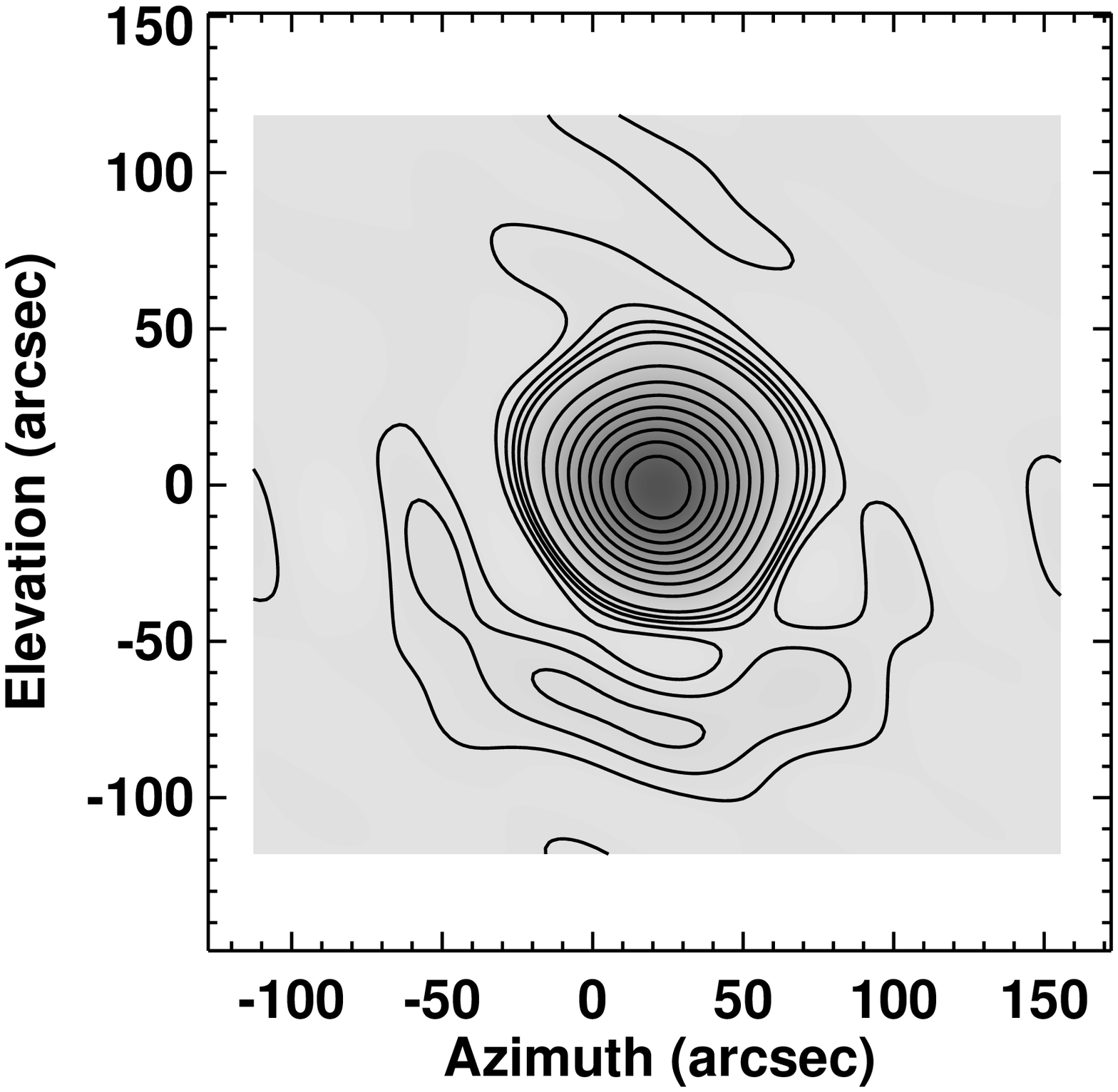}
   \includegraphics[height=5.3cm]{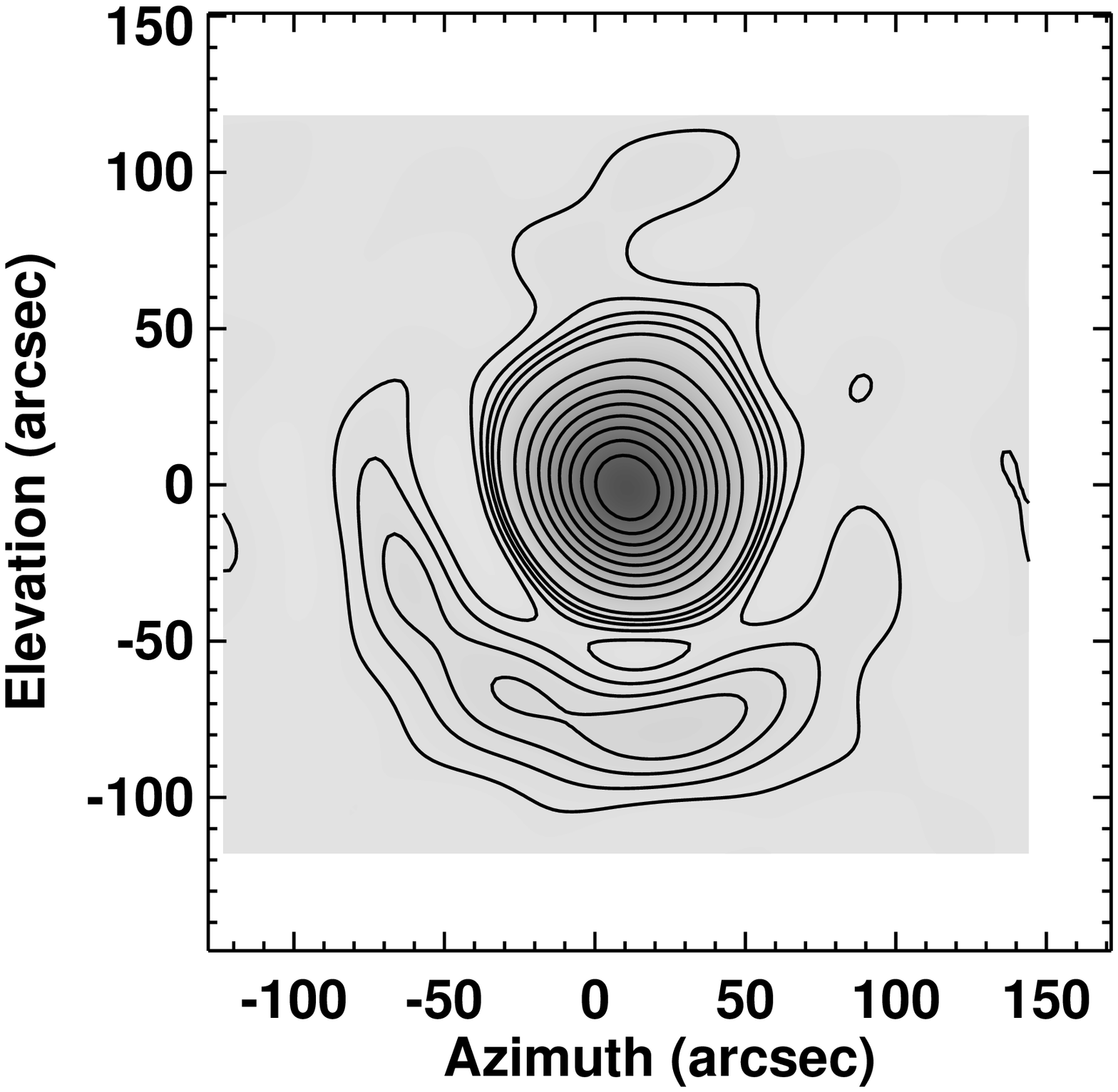}
   \end{tabular}
   \end{center}
   \caption[bpfix]
   {\label{fig:bpfix}
    Beam pattern maps of the SMA antenna No.~4 at various
    elevation angles with a fixed subreflector position optimized
    at the elevation angle of $\sim19^{\circ}$.
    The beam pattern maps were taken toward Jupiter on December 16th,
    2000 for the left figures, and on December 9th, 2000 for
    the others.
    The left, middle, and right figures are taken at the elevation
    angle of $35^{\circ}$, $65^{\circ}$, and $82^{\circ}$,
    respectively.
    Top row maps are observed images, and the bottom row maps are
    deconvolved images.
    The observed frequency was 237~GHz at the local oscillator
    frequency.
    Contour levels are $1, 3, 5, 7, 10, 20, 30, \cdots, 90\%$ of
    the peak.
   }
   \end{figure}

To see the degree of gravitational deformation, we mapped the beam
patterns of the SMA antennas.
The beam pattern maps were obtained with azimuthal or elevational
scans across bright astronomical sources, such as planets.
The data taking rate and the scan strip spacing were one-third of
the Nyquist sampling of the SMA primary beam size ($52''$ at
230~GHz).
The beam pattern maps were obtained at various antenna elevation
angles with a fixed subreflector position optimized at the elevation
angle of $\sim19^{\circ}$.
The examples of the beam pattern maps for the antenna No.~4 are
shown in the top row of Fig.~\ref{fig:bpfix}.
The top-left map was taken on December 16th, 2000, and the
top-middle and top-right maps were on December 9th, 2000.
These example maps were taken toward Jupiter with azimuthal scans
and with the observing frequency of 237~GHz at local oscillator
frequency (232~GHz and 242~GHz for the LSB and USB frequencies,
respectively, with the bandwidth of 2~GHz for each sideband).
The maps were obtained using a continuum detector, which does not
have any sideband separation system, so that the maps have a 4~GHz
effective bandwidth.

The bottom row of Fig.~\ref{fig:bpfix} shows the deconvolved beam
pattern maps (deconvolved with the size of Jupiter) of the top row
figures.
We used the Jupiter size of $48.4''\times45.4''$ and
$48.0''\times45.1''$ with the position angle of $100^{\circ}$
clockwise from azimuth direction for the measurement data of December
9th and 16th, 2000, respectively.
We assume the intensity distribution of Jupiter as a uniform disk.
At a low elevation angle of $35^{\circ}$ (the bottom-left map of
Fig.~\ref{fig:bpfix}), there are sidelobes with at most a few \%
level in the beam pattern map.
However, as the elevation angle increases, a lopsided sidelobe,
which locates below the main beam, increases significantly.
The bottom-middle and the bottom-right maps of Fig.~\ref{fig:bpfix}
are taken at the elevation angles of $65^{\circ}$ and $82^{\circ}$,
respectively, and the peak of the lopsided sidelobe levels are about
$\sim5\%$ and $\sim8\%$, respectively.
The peak of the lopsided sidelobe level rises linearly with the
increase of the offset of the antenna elevation angle from the
subreflector optimized elevation angle of $\sim19^{\circ}$, and the
peak of the sidelobe increases about 1\% in every $10^{\circ}$ change
of the elevation angle.

%-------------
%-------------

%%-----------------------------------------------------------
\subsection{Antenna Efficiencies with Fixed Focus Position}
\label{sect:efffix}

In addition to the beam pattern deformation, the gravitational
deformation also causes a decrease of antenna efficiencies.
We also performed antenna efficiency measurements with observing
targets with known temperature, namely planets (Jupiter for these
measurements), an ambient load (antenna cabin temperature), and
a cold load (liquid nitrogen).
The measurement was done on November 17th, 2001 with the SMA antenna
No.~7.
The measurement frequency was the same as the beam pattern mapping
mentioned above.
The subreflector position was again fixed at the optimized position
at the elevation angle of $\sim19^{\circ}$.

\begin{figure}
   \begin{center}
   \begin{tabular}{c}
   \includegraphics[height=7cm]{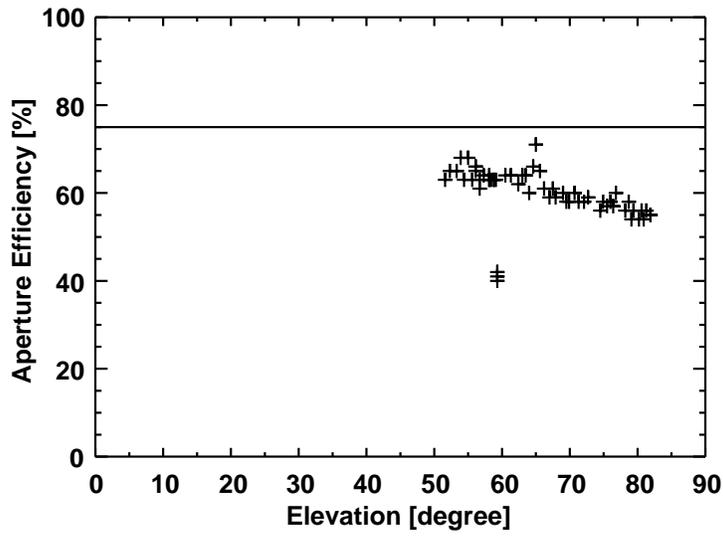}
   \end{tabular}
   \end{center}
   \caption[efffix]
   {\label{fig:efffix}
    Aperture efficiency measurement results for the SMA antenna No.~7
    at various elevation angles with a fixed subreflector position
    optimized at the elevation angle of $\sim19^{\circ}$.
    The measurement was done toward Jupiter on November 17th, 2001,
    and the observed frequency was 237~GHz at the local oscillator
    frequency.
   }
   \end{figure}

Fig.~\ref{fig:efffix} plots the measurement results of aperture
efficiencies as a function of antenna elevation angles.
The aperture efficiency with an ideal focus (subreflector) position
is about 0.75 around 237~GHz, which is also shown in the figure with
a straight solid line.
The aperture efficiency of 0.75 around 237~GHz is calculated as
follows:  The aperture efficiency can be described as
\begin{equation}
\label{eq:eta}
\eta_{\rm a} = \eta_{0}e^{-(4\pi\epsilon/\lambda)^2},
%\eta_{\rm a} = \eta_{0}\exp
%  \left{-\left(\frac{4\pi\epsilon}{\lambda}\right)^{2}\right},
\end{equation}
where $\eta_{\rm a}$ is an actual aperture efficiency, $\eta_{0}$ is
an ideal antenna efficiency, $\epsilon$ is a total antenna
surface/optics error, and $\lambda$ is an observing wavelength
\cite{ruz52,but03}.
The ideal aperture efficiency for the SMA antennas is 0.805
\cite{pai95}, and the total SMA antenna surface/optics error at that
time was around 27~$\mu$m.
Substitute these numbers with the measurement wavelength of 1.27~mm
($\sim237$~GHz), the actual aperture efficiency can be calculated as
0.75.
As you can see in the figure, the aperture efficiency at lower
elevation angle is close to the focus optimized efficiency of 0.75
but still lower than this value, and it gradually decreases with the
elevation angle away from the optimized elevation angle of
$\sim19^{\circ}$.

%-------------
%-------------

%% This table is carefully placed in the source file to make
%% it appear at bottom of page, but above the footnotes.
%% Use of [h] in following command forces table to appear "here".
\begin{table}[b]
\caption{Elevation angle dependence of 237~GHz aperture and main
  beam efficiencies at each $10^{\circ}$ elevation angle bin.
  The data are the same as that shown in Fig.~\ref{fig:efffix}.
  The errors in the efficiency values are $1\sigma$ standard
  deviations of the data.
  The numbers of data indicate the data points within an elevation
  angle bin.}
\label{tab:efffix}
\begin{center}       
\begin{tabular}{|c|c|c|c|} %% this creates four columns
%% |l|l| to left justify each column entry
%% |c|c| to center each column entry
%% use of \rule[]{}{} below opens up each row
\hline
\rule[-1ex]{0pt}{3.5ex} Elevation & Aperture Eff. & Main Beam Eff. & No.\ of data \\
\rule[-1ex]{0pt}{3.5ex} (degrees) &     (\%)      &      (\%)      &              \\
\hline
\rule[-1ex]{0pt}{3.5ex}  $50-60$  &   $64\pm2$    &    $75\pm2$    &      20      \\
\rule[-1ex]{0pt}{3.5ex}  $60-70$  &   $62\pm3$    &    $73\pm4$    &      16      \\
\rule[-1ex]{0pt}{3.5ex}  $70-80$  &   $58\pm2$    &    $68\pm2$    &      14      \\
\rule[-1ex]{0pt}{3.5ex}  $80-82$  &   $55\pm1$    &    $64\pm1$    & \phantom{1}6 \\
\hline
\end{tabular}
\end{center}
\end{table} 

Table~\ref{tab:efffix} summarizes the measurement results of aperture
and main beam efficiencies at each $10^{\circ}$ elevation angle bin.
The errors for the efficiencies in the table indicate the $1\sigma$
standard deviations of the data in each bin.
As you can see from the figure and the table, the efficiencies
decrease as the elevation angle offset increases from the
subreflector optimized elevation angle of $\sim19^{\circ}$.
The decrement of the efficiencies is about a few \% in every
$10^{\circ}$ increase of the elevation angle.

%%%%%%%%%%%%%%%%%%%%%%%%%%%%%%%%%%%%%%%%%%%%%%%%%%%%%%%%%%%%%
\section{FOCUS POSITION OPTIMIZATION}
\label{sect:opt}

Previous section shows that without optimizing the subreflector
positions, the antenna characteristics degrade significantly because
of the gravity.
We therefore needed to find out a way to avoid the degradation.

As mentioned above (Sect.~\ref{sect:intro}), there are two possible
reasons for the antenna elevation angle dependences of the SMA
antenna beam patterns and antenna efficiencies; one is the
gravitational deformation of the main reflector, and the other is the
gravitational sag of the subreflector.
The gravitational deformation of the main reflector cannot be
compensated in real-time, since there are no motors to move the main
reflector panels remotely on the SMA antennas.
On the other hand, it is possible to move the subreflector remotely
in four directions; X (azimuth axis), Y (elevation axis), Z (radio
optical axis), and tilt (azimuth axis) directions.
It is therefore possible to change the focus positions in real-time,
to compensate the effect of the gravitational sag of the
subreflector, and to use the focus position that best fits the
deformed primary reflector.
We therefore measured the optimized subreflector positions, namely
focus positions, at various elevation angles for all the SMA
antennas (No.~1 -- 8).

The measurements were performed toward bright and known structure
sources, namely planets (mostly Jupiter, Saturn, and Venus), between
2002 -- 2004.
The sources were observed at five different subreflector positions
in each direction (X, Y, and Z axes) with the point-to-point
separation as the Nyquist sampling of the SMA primary beam size
($52''$ at 230~GHz).
The five output amplitude values of each axis were fitted with
a 2nd order polynomial, and the fitted peak value was assume to be
the optimized position of each axis.
A measurement of each axis took several minutes, and was done by
turns.
Since this measurement was sensitive to the antenna pointing,
single-dish pointings were done every 1 hour or so.
We did not optimize the subreflector tilt direction.
This is because the tilt direction can be compensate with
optimizing the X axis \cite{sai06}.

\begin{figure}
   \begin{center}
   \includegraphics{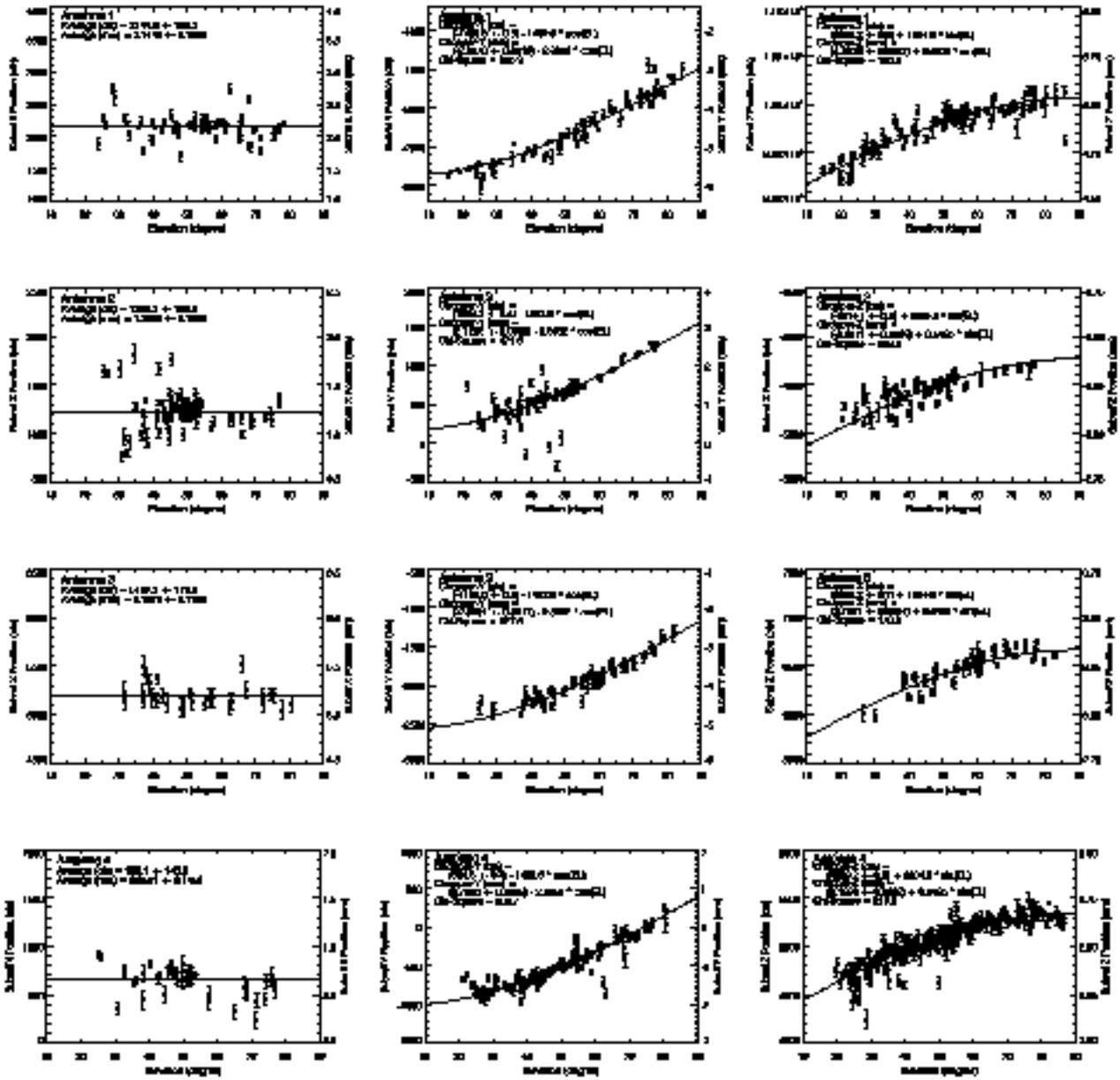}
%   \begin{tabular}{ccc}
%   \vspace{-2.3cm}
%   \includegraphics[height=6.4cm]{subref-pos_a1x_rx230.eps}
%   \hspace{1.1cm}
%   \includegraphics[height=6.4cm]{subref-pos_a1y_rx230.eps}
%   \hspace{1.1cm}
%   \includegraphics[height=6.4cm]{subref-pos_a1z_rx230.eps}\\
%   \vspace{-2.3cm}
%   \includegraphics[height=6.4cm]{subref-pos_a2x_rx230.eps}
%   \hspace{1.1cm}
%   \includegraphics[height=6.4cm]{subref-pos_a2y_rx230.eps}
%   \hspace{1.1cm}
%   \includegraphics[height=6.4cm]{subref-pos_a2z_rx230.eps}\\
%   \vspace{-2.3cm}
%   \includegraphics[height=6.4cm]{subref-pos_a3x_rx230.eps}
%   \hspace{1.1cm}
%   \includegraphics[height=6.4cm]{subref-pos_a3y_rx230.eps}
%   \hspace{1.1cm}
%   \includegraphics[height=6.4cm]{subref-pos_a3z_rx230.eps}\\
%%   \vspace{-2.3cm}
%   \includegraphics[height=6.4cm]{subref-pos_a4x_rx230.eps}
%   \hspace{1.1cm}
%   \includegraphics[height=6.4cm]{subref-pos_a4y_rx230.eps}
%   \hspace{1.1cm}
%   \includegraphics[height=6.4cm]{subref-pos_a4z_rx230.eps}
%   \end{tabular}
   \end{center}
   \caption[sropt1]
   {\label{fig:sropt1}
    Antenna elevation angle dependences of the optimized subreflector
    positions for all the SMA antennas (No.~1 -- 8).
    The first, second, and third columns are the subreflector X
    (azimuth), Y (elevation), and Z (radio optics) axes results,
    respectively.
    The left-hand side vertical axes indicate the subreflector
    positions in counts, and right-hand side axes indicate in
    millimeters.
    The straight solid lines of the X axis are the average values
    of the data points for each antenna, and the curved lines of
    the Y and Z axes are the fitted results with fixed coefficients
    and free zero-point offsets.
    The fixed coefficients are the average values of the coefficients
    of all the antennas.
   }
   \end{figure}

The data points in Fig.~\ref{fig:sropt1} show the measured optimized
subreflector positions in each axis for all the SMA antennas.
All the optimized subreflector (focus) position data plots basically
show the same behavior; optimized positions for the X axis (the first
column of Fig.~\ref{fig:sropt1}) do not depend on the antenna
elevation angle, but Y and Z axes (the second and the third columns
of Fig.~\ref{fig:sropt1}, respectively) obviously depend on it.
The non-dependence of the X axis on the elevation angle can easily be
explained, since the X axis is parallel to the elevation axis and
therefore should not be affected by changes in elevation angle.
%perpendicular to the antenna elevation
%axis and should not change with the elevation angle.

We try to fit a simple function to the elevation angle dependence
of the Y axis of the subreflector position (the second column of
Fig.~\ref{fig:sropt1}).
We assume a function
\begin{equation}
\label{eq:y}
Y = B - A\cos(EL),
\end{equation}
where $Y$ is the measured optimized subreflector Y axis positions,
$B$ is a zero-point offset, $A$ is a coefficient, and $EL$ is
the elevation angle of an antenna.
We then fit this function to the data points, and we can fit the data
very well.
The first term $B$ and the second term $-A\cos(EL)$ can be
explained as a constant term (DC offset) and a compensation term for
the subreflector sag caused by the gravity, respectively.

We also try to fit a simple function to the elevation angle
dependence of the Z axis of the subreflector position (the third
column of Fig.~\ref{fig:sropt1}).
We again assume a function
\begin{equation}
\label{eq:z}
Z = D + C\sin(EL),
\end{equation}
where $Z$ is the measured optimized subreflector Z axis positions,
$D$ is a zero-point offset, $C$ is a coefficient, and $EL$ is
the elevation angle of an antenna.
We fit this function to the data points, and again we can fit
the data very well.
The first term $D$ can be explained as a constant term (DC offset).
The second term $C\sin(EL)$ can be understood as follows:
When the elevation angle is high, the edge of the primary dish may
sag due to the gravity, and the primary dish may deform (open) toward
a larger paraboloid \cite{raf91}.
This deformation causes the focus far from the usual, and
the optimized subreflector Z axis positions will go to larger values.

%-------------
%-------------

%-------------
   \addtocounter{figure}{-1}
   \begin{figure}
   \begin{center}
   \includegraphics{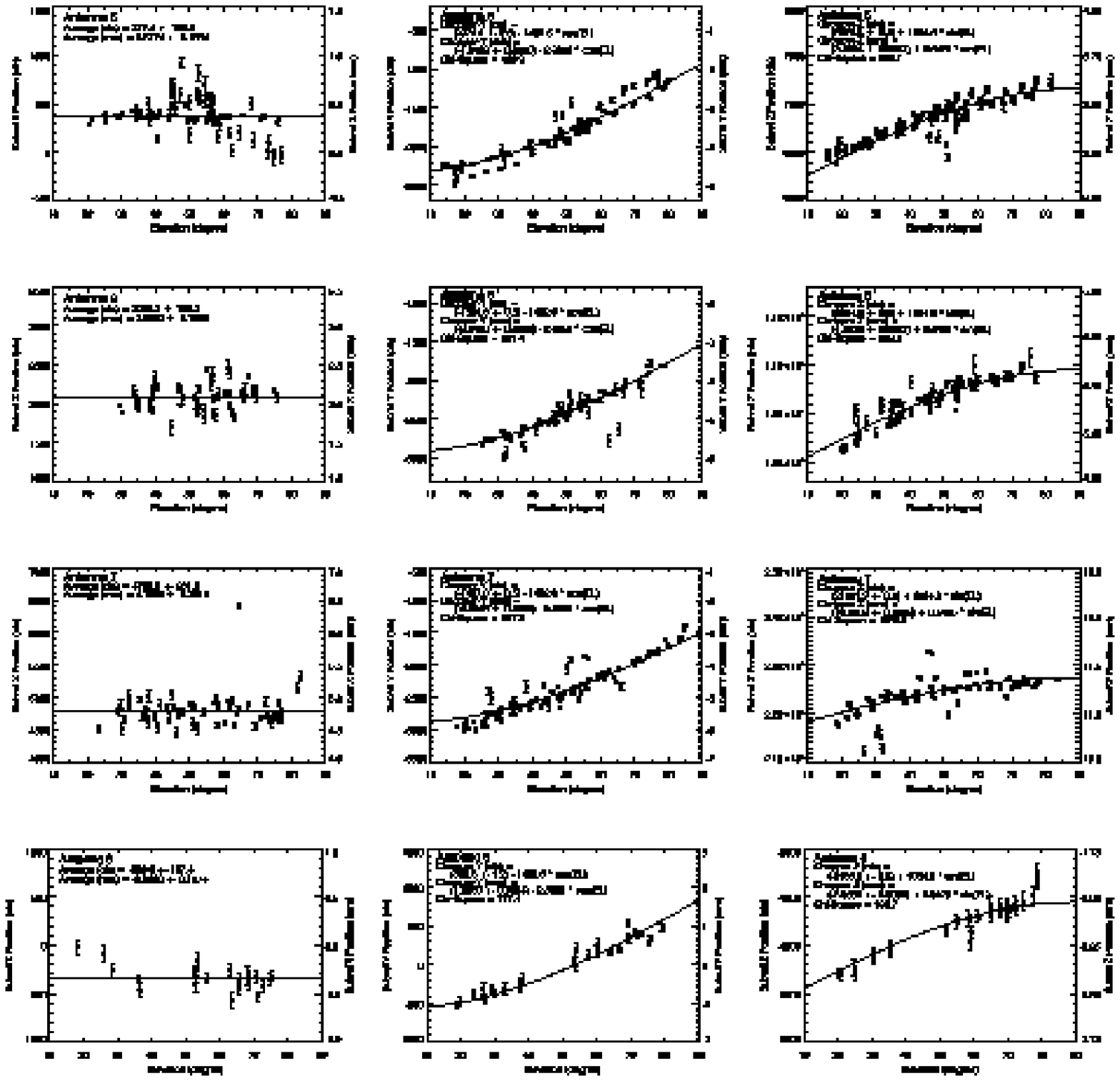}
%   \begin{tabular}{ccc}
%   \vspace{-2.3cm}
%   \includegraphics[height=6.4cm]{subref-pos_a5x_rx230.eps}
%   \hspace{1.1cm}
%   \includegraphics[height=6.4cm]{subref-pos_a5y_rx230.eps}
%   \hspace{1.1cm}
%   \includegraphics[height=6.4cm]{subref-pos_a5z_rx230.eps}\\
%   \vspace{-2.3cm}
%   \includegraphics[height=6.4cm]{subref-pos_a6x_rx230.eps}
%   \hspace{1.1cm}
%   \includegraphics[height=6.4cm]{subref-pos_a6y_rx230.eps}
%   \hspace{1.1cm}
%   \includegraphics[height=6.4cm]{subref-pos_a6z_rx230.eps}\\
%   \vspace{-2.3cm}
%   \includegraphics[height=6.4cm]{subref-pos_a7x_rx230.eps}
%   \hspace{1.1cm}
%   \includegraphics[height=6.4cm]{subref-pos_a7y_rx230.eps}
%   \hspace{1.1cm}
%   \includegraphics[height=6.4cm]{subref-pos_a7z_rx230.eps}\\
%%   \vspace{-2.3cm}
%   \includegraphics[height=6.4cm]{subref-pos_a8x_rx230.eps}
%   \hspace{1.1cm}
%   \includegraphics[height=6.4cm]{subref-pos_a8y_rx230.eps}
%   \hspace{1.1cm}
%   \includegraphics[height=6.4cm]{subref-pos_a8z_rx230.eps}
%   \end{tabular}
   \end{center}
   \caption[sropt2]
   {\label{fig:sropt2}
    \it Continue.
   }
   \end{figure}
%-------------

The best focus positions for the Y and Z axes of the subreflector
move about 2.5~mm and 0.43~mm as the elevation angle changes from
$12^{\circ}$ to $85^{\circ}$  (this range corresponds to the lower
and upper elevation angle limits of the SMA antennas).
These values are similar to or larger than the SMA operating
frequencies of 180~GHz (1.7~mm) to 900~GHz (0.33~mm), so that
observe astronomical sources without optimizing the subreflector
positions will degrade the data quality significantly.

%%%%%%%%%%%%%%%%%%%%%%%%%%%%%%%%%%%%%%%%%%%%%%%%%%%%%%%%%%%%%
\section{MEASUREMENT RESULTS WITH OPTIMIZED FOCUS POSITION}
\label{sect:measopt}

%%-----------------------------------------------------------
\subsection{Beam Pattern Maps with Optimized Focus Position}
\label{sect:bpopt}

\begin{figure}
   \begin{center}
   \begin{tabular}{cc}
   \includegraphics[height=6.6cm]{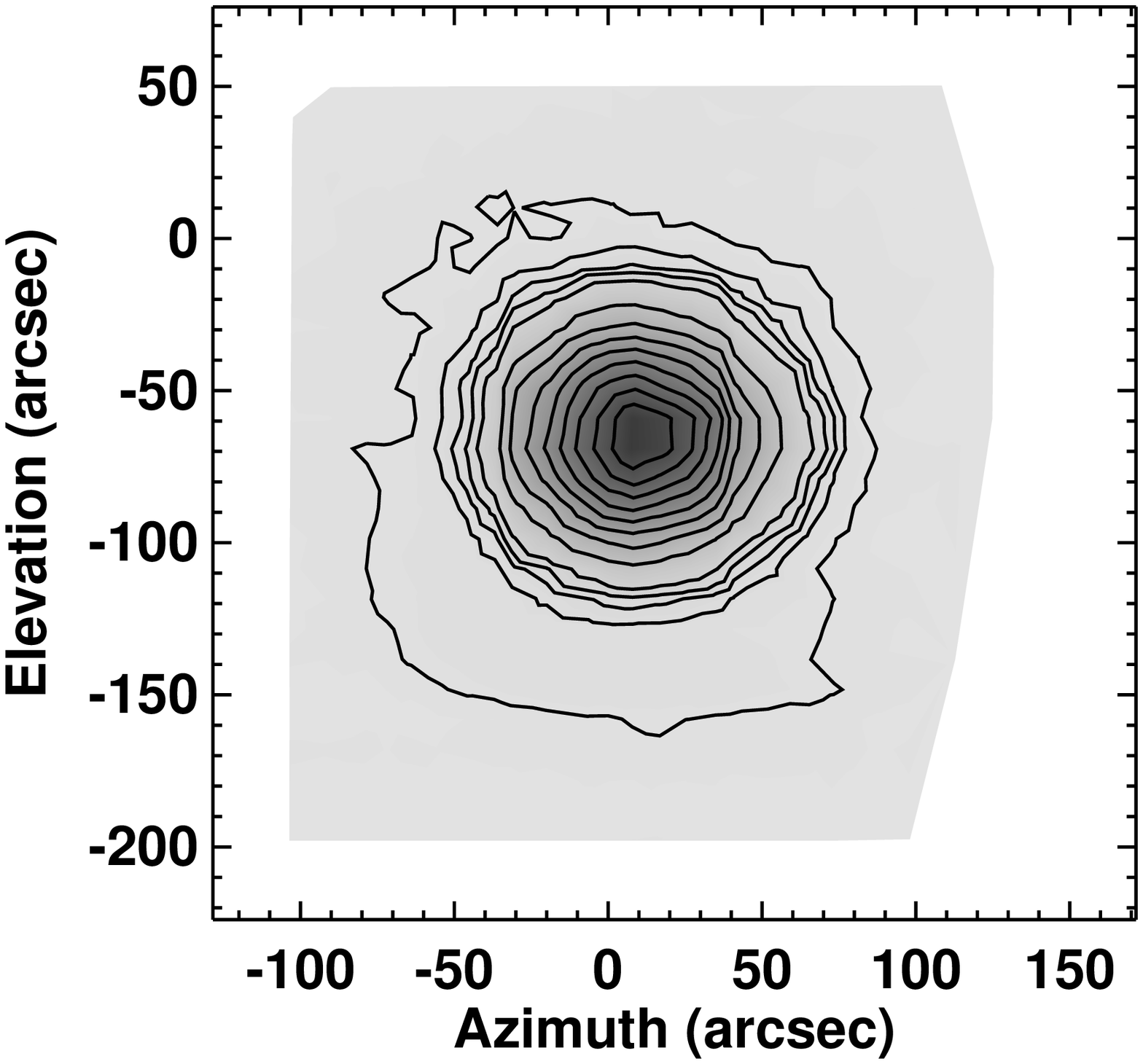}
   \includegraphics[height=6.6cm]{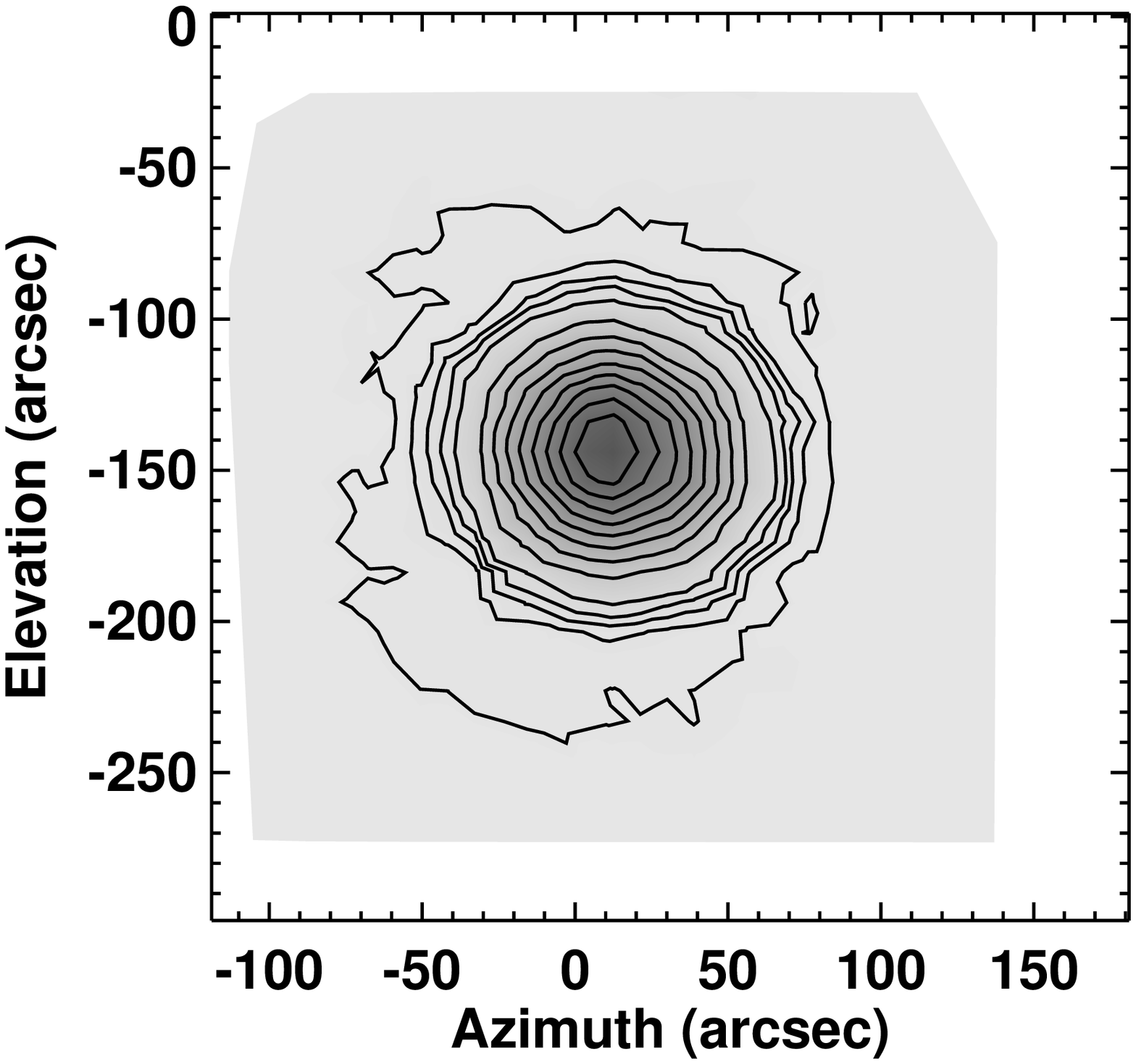}\\
   \includegraphics[height=6.9cm]{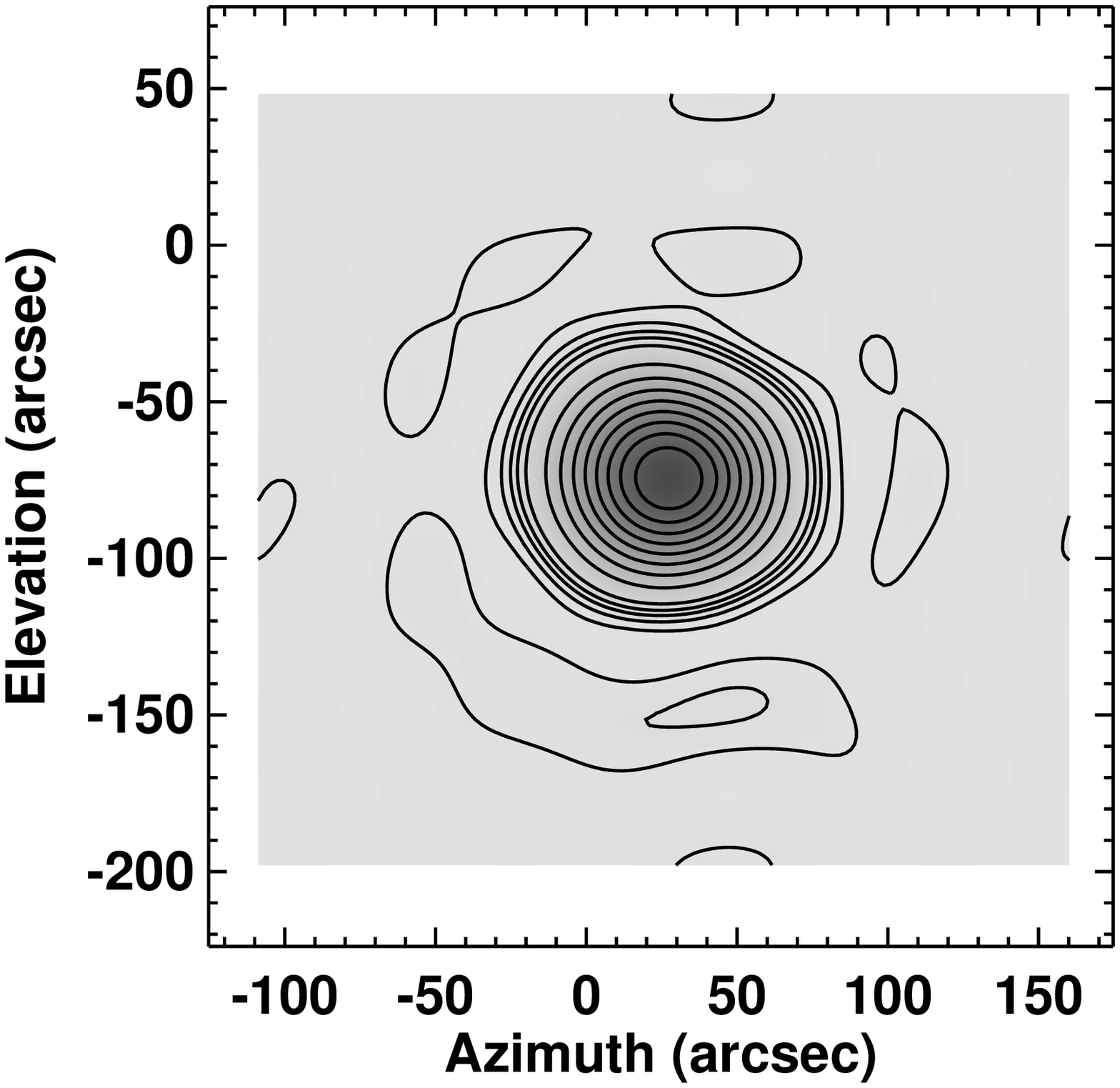}
   \includegraphics[height=6.9cm]{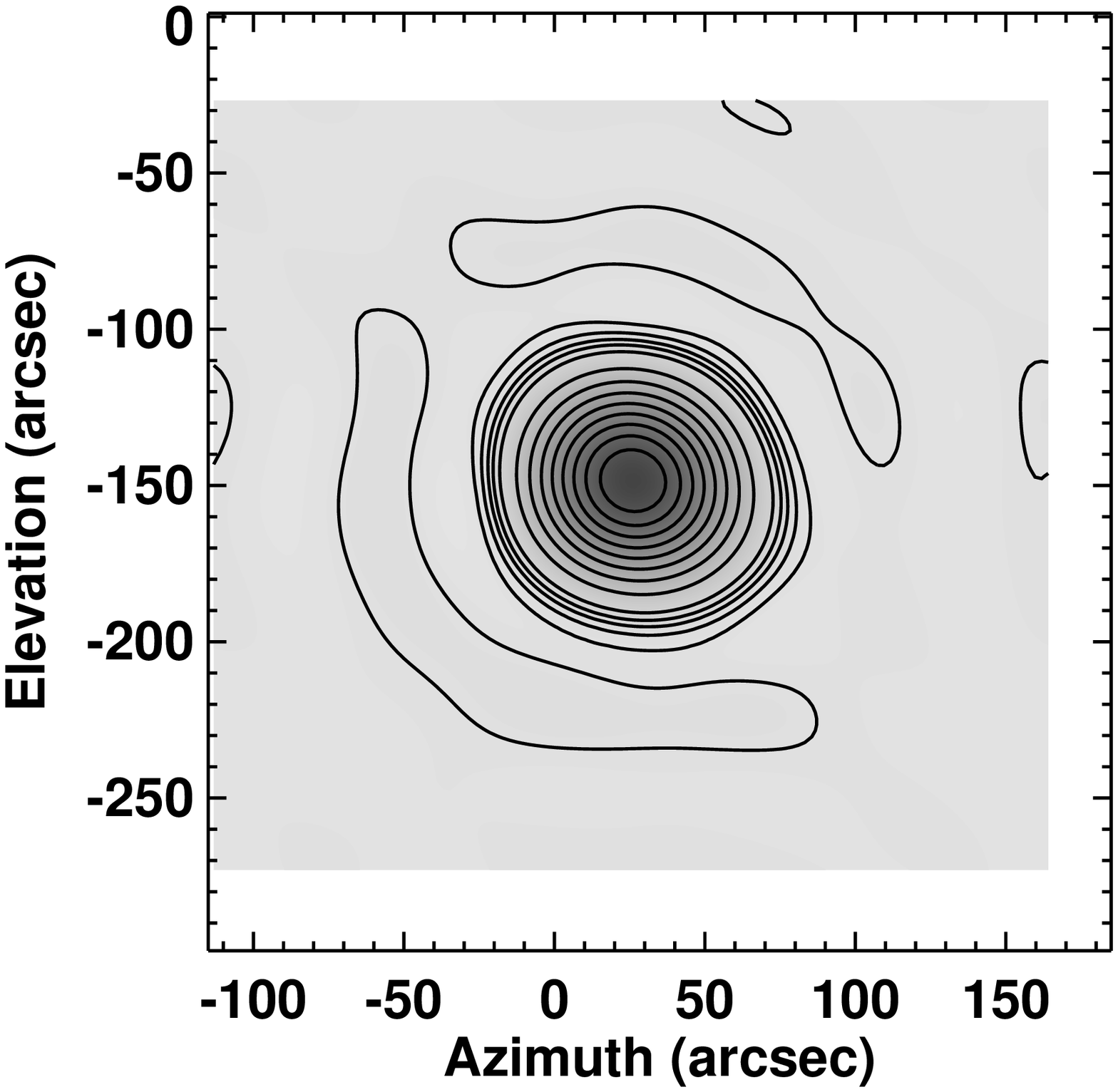}
   \end{tabular}
   \end{center}
   \caption[bpopt]
   {\label{fig:bpopt}
    Beam pattern maps of the SMA antenna No.~4 at various
    elevation angles with subreflector positions optimized at the
    measured elevations.
    The beam pattern maps were taken toward Jupiter on December 16th,
    2000.
    The left and right figures are taken at the elevation angle of
    $60^{\circ}$ and $75^{\circ}$, respectively.
    Top row maps are observed images, and the bottom row maps are
    deconvolved images.
    The observed frequency was 237~GHz at the local oscillator
    frequency.
    Contour levels are $1, 3, 5, 7, 10, 20, 30, \cdots, 90\%$ of
    the peak.
   }
   \end{figure}

Using the focus curves obtained in the previous section, we performed
beam pattern mapping at various elevation angles with optimizing
subreflector positions using the SMA antenna No.~4.
The top row of Fig.~\ref{fig:bpopt} shows the beam pattern maps
taken toward Jupiter on December 16th, 2000.
The elevation angles were $60^{\circ}$ (left column) and $75^{\circ}$
(right column), respectively.
Other measurement settings are the same as the measurements mentioned
in Sect~\ref{sect:bpfix}.

%-------------
%-------------

The bottom row of Fig.~\ref{fig:bpopt} show the deconvolved beam
pattern maps (deconvolved with the size of Jupiter) of the top row
figures.
The parameters of Jupiter for the deconvolution is the same as that
used in Sect.~\ref{sect:bpfix}.
Both of the maps show symmetric beam patterns and low sidelobe levels
of at most a few \%.
Compared with the defocused beam pattern images
(Fig.~\ref{fig:bpfix}), the focus-optimized beam pattern images do
not have the lopsided sidelobes, even at high antenna elevation
angles.
The positions and the intensity levels of these low-level circular
sidelobes in the deconvolved beam pattern maps are consistent with
those of the first sidelobe of the SMA antenna derived with
theoretical SMA beam pattern calculations \cite{pai06}.

%%-----------------------------------------------------------
\subsection{Aperture Efficiencies with Optimized Focus Position}
\label{sect:effopt}

We also performed the aperture efficiency measurements with optimized
focus positions.
The measurements were done on February 9th and 10th, 2005, with
observing targets with known temperature, namely planets (Jupiter
and Saturn for these measurements), an ambient load (antenna cabin
temperature), and a cold load (liquid nitrogen).
The SMA antenna No.~1--4 were used for the measurements, and the
measurement frequencies were 230~GHz for February 9th and 345~GHz
for February 10th.
The subreflector positions were optimized at the measurement
elevation angle based on the focus curves obtained in the previous
section.

\begin{figure}
   \begin{center}
   \begin{tabular}{cc}
   \includegraphics[height=6cm]{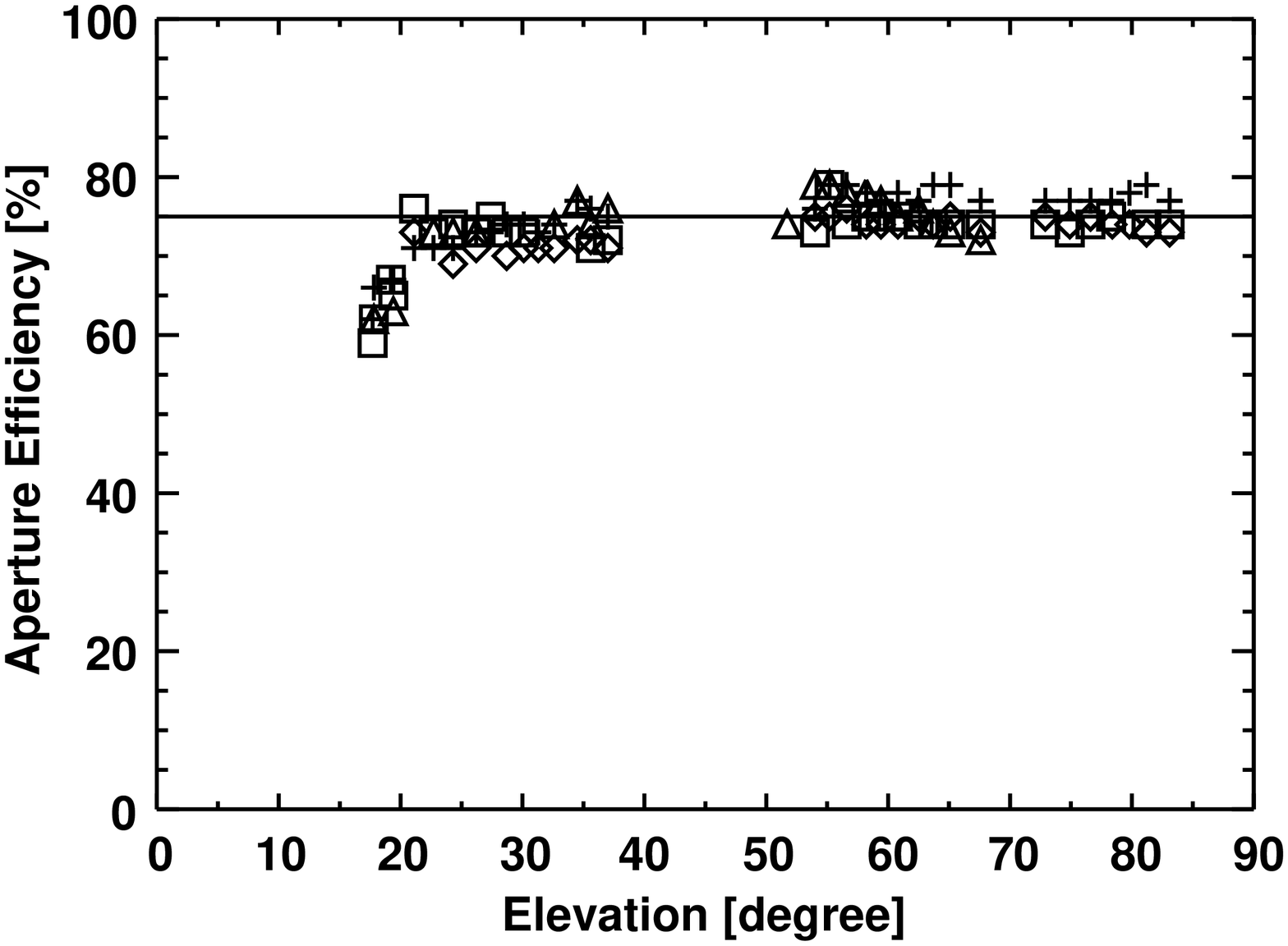}
   \hspace{0.3cm}
   \includegraphics[height=6cm]{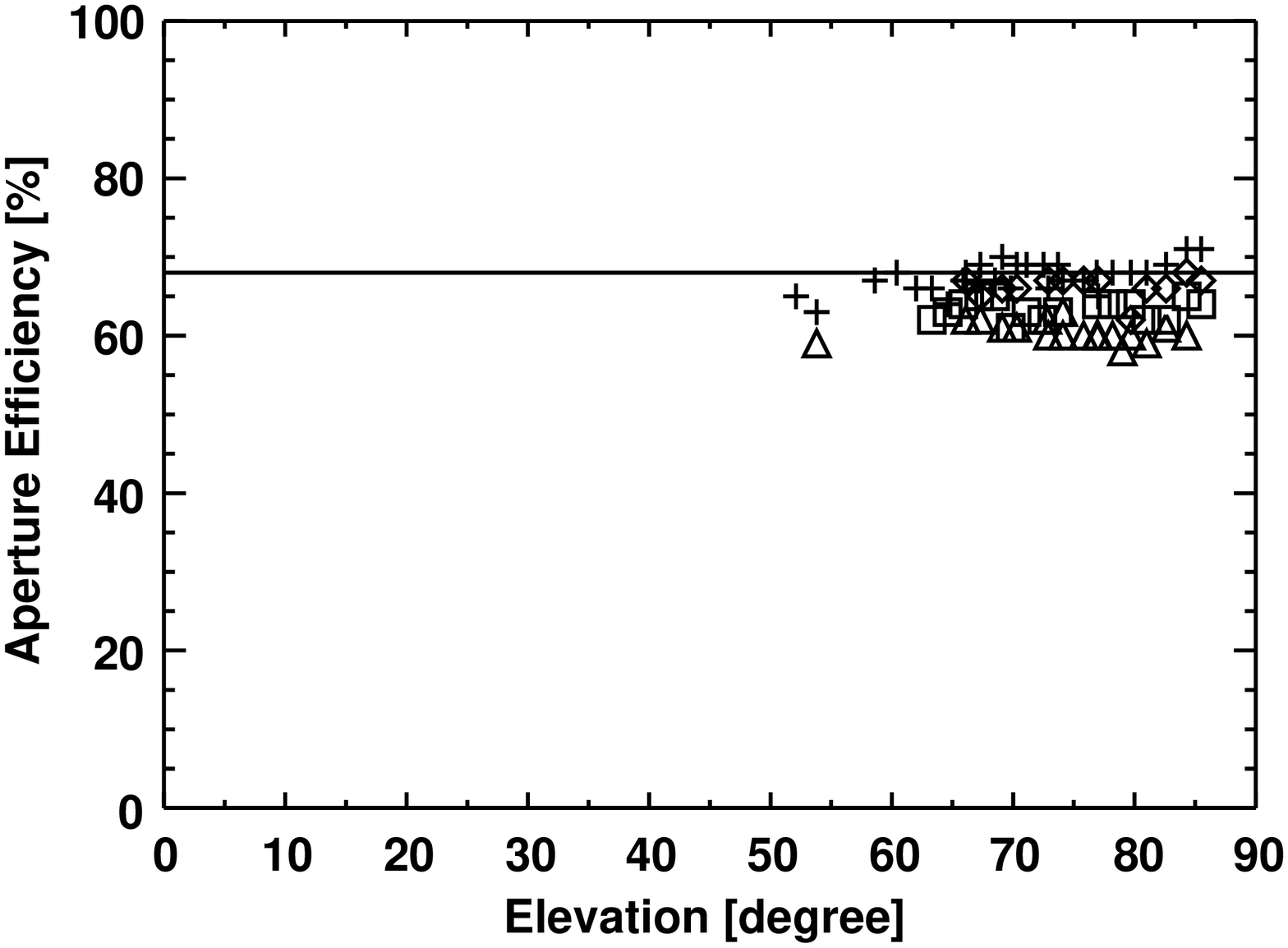}
   \end{tabular}
   \end{center}
   \caption[effopt]
   {\label{fig:effopt}
    Aperture efficiency measurement results for the SMA antennas at
    various elevation angles with a optimized subreflector (focus)
    positions using the focus curves derived in Sect.~\ref{sect:opt}.
    The cross, diamond, triangle, and square marks indicate the
    measurement results for the SMA antennas No.~1, 2, 3, and 4,
    respectively.
    The left and right figures indicate the 230~GHz and 345~GHz
    measurement results, respectively.
    The measurement was done toward Jupiter and Saturn on February
    9th, 2005 for the 230~GHz measurements and 10th for the 345~GHz
    measurements.
   }
   \end{figure}

The measurement results are shown in Fig.~\ref{fig:effopt}.
Left plot is the measurements at 230~GHz, and the right plot is at
345~GHz.
The straight solid line in each plot indicates the aperture
efficiency with an ideal focus position calculated using
Eq.~(\ref{eq:eta}) with the same assumption as mentioned at
Sect.~\ref{sect:efffix} with different frequency for each plot, which
is 0.75 for the left plot (230~GHz) and 0.68 for the right plot
(345~GHz).
As you can see the efficiencies in both frequencies are almost
constant at the focus optimized aperture efficiencies in almost
all elevation angles with all measured antennas.
The decrease at the low elevation angle (lower than $20^{\circ}$) is
not clear, possibly due to atmosphere, but no repeated measurements
so far.

%-------------
%-------------

%%%%%%%%%%%%%%%%%%%%%%%%%%%%%%%%%%%%%%%%%%%%%%%%%%%%%%%%%%%%%
\section{REAL-TIME FOCUS OPTIMIZATION}
\label{sect:real}

The disappearance of the lopsided sidelobes in beam pattern maps
and the constant aperture efficiencies at most of the elevation
with optimizing the subreflector (focus) positions shown in the
previous section indicate that most of the gravitational deformation
of the SMA antennas can be compensated with optimizing the
subreflector positions.
To obtain constant antenna characteristics at any given antenna
elevation angle, namely at the actual scientific observations, the
real-time subreflector position optimization is needed.
However, there are several points to concern for the real-time
subreflector position optimization with the SMA (and also for other
interferometers).
First, the gravity term coefficients $A$ and $C$ in Eq.~(\ref{eq:y})
and (\ref{eq:z}) have to be the same for all the antennas.
If one used the different coefficients for each antenna, the gravity
compensation terms (the second terms in Eq.~(\ref{eq:y}) and
(\ref{eq:z})) will induce additional delay changes, and complicates
the delay calculations of the SMA.
Second, the zero-point (DC) offsets $B$ and $D$ is better to use
optimized values for each antenna.
These terms do not induce any delay changes or other defects on
the SMA, so using the optimized values will give the best performance
of the antennas.
Note that if receiver components, especially SIS chips or horns, have
been changed, it is better to re-measure the zero-points offsets.
Third, changing the subreflector positions induces the change in
the antenna pointing.
In case of the SMA antennas, 1~mm change in X and Y axes introduce
$66''.3$ offset in the sky plane.
Therefore, a real-time pointing model correction is also needed.
Since the subreflector positions move along the elevation direction,
we only need to introduce the pointing model correction for the
elevation direction.

To do the real-time subreflector position optimization, we need to
find out the common gravity term coefficients.
We fitted the functions Eq.~(\ref{eq:y}) and (\ref{eq:z}) to
the measured data points for all the SMA antennas, and calculated
the average gravity term coefficients $A$ and $C$ as the common
coefficients.
The average values turned to be 2.8052 and 0.5425 in millimeters
(1402.6 and 1084.9 in counts) for the coefficients $A$ and $C$,
respectively.
We then fitted the functions Eq.~(\ref{eq:y}) and (\ref{eq:z})
to the measured data points using the common coefficients derived
above (only the zero-point offsets are the free parameters), and
the results are shown in Fig.~\ref{fig:sropt1} (solid lines).
As you can see, all the measured data points can be fitted well
(well within the scatters of the data points) with the common
coefficients.

Currently, the SMA is operating all the observations with the
real-time subreflector (focus) position optimization using these
fitted focus curves in Fig~\ref{fig:sropt1} with applying all the
concerns mentioned above.
Measurement of aperture efficiencies and beam patterns with
real-time subreflector position optimization show constant
efficiencies and low sidelobe levels at all the elevation angles.
This strongly indicates the success of the real-time subreflector
position optimization.

%%%%%%%%%%%%%%%%%%%%%%%%%%%%%%%%%%%%%%%%%%%%%%%%%%%%%%%%%%%%%
\acknowledgments     %>>>> equivalent to \section*{ACKNOWLEDGMENTS}

We would like to thank Scott Paine for valuable comments.
Thanks are also due to all the past and present SMA staffs, who
helped to measure beam patterns, antenna efficiencies, and focus
positions.
The Submillimeter Array is a joint project between the Smithsonian
Astrophysical Observatory and the Academia Sinica Institute of
Astronomy and Astrophysics and is funded by the Smithsonian
Institution and the Academia Sinica.

%%%%%%%%%%%%%%%%%%%%%%%%%%%%%%%%%%%%%%%%%%%%%%%%%%%%%%%%%%%%%
%%%%% References %%%%%

\end{document}